\documentclass[preprint,10pt]{elsarticle}

\usepackage{amssymb}
\usepackage{amsmath}

\newtheorem{thm}{Theorem}[section]

\makeatletter

\@addtoreset{equation}{section}

\def\be{{\beta}}

\def\de{{\delta}}
\def\ep{{\varepsilon}}

\def\si{{\sigma}}
\def\th{{\theta}}
\def\bbe{{\text{\boldmath $\beta$}}}

\def\bphi{{\text{\boldmath $\phi$}}}

\def\muh{{\widehat \mu}}

\def\bbeh{{\widehat \bbe}}

\def\mut{\widetilde{\mu}}
\def\pt{\widetilde{p}}
\def\Si{{\Sigma}}

\def\bSi{{\text{\boldmath $\Si$}}}
\def\0{{\text{\boldmath $0$}}}
\def\1{{\text{\boldmath $1$}}}
\def\c{{\text{\boldmath $c$}}}
\def\s{{\text{\boldmath $s$}}}
\def\u{{\text{\boldmath $u$}}}
\def\x{{\text{\boldmath $x$}}}
\def\y{{\text{\boldmath $y$}}}
\def\v{{\text{\boldmath $v$}}}
\def\I{{\text{\boldmath $I$}}}
\def\A{{\text{\boldmath $A$}}}

\def\J{{\text{\boldmath $J$}}}
\def\X{{\text{\boldmath $X$}}}

\def\Z{{\text{\boldmath $Z$}}}

\def\diag{{\rm diag}}

\def\barx{\bar{\x}}
\def\barX{\bar{\X}}
\def\bary{\bar{y}}
\def\barY{\bar{Y}}
\def\barep{\bar{\ep}}
\def\Re{\mathbb{R}}
\def\Var{{\rm Var}}

\def\E{{\rm E}}
\begin{document}

\begin{frontmatter}



\title{Bayesian Estimators in Uncertain Nested Error Regression Models}


\author{Shonosuke Sugasawa} 
\ead{shonosuke622@gmail.com}
\address{The Institute of Statistical Mathematics, 10-3 Midori-cho, Tachikawa-shi, Tokyo 190-8562, Japan.}

\author{Tatsuya Kubokawa}
\ead{tatsuya@e.u-tokyo.ac.jp}
\address{Faculty of Economics, University of Tokyo, 7-3-1 Hongo, Bunkyo-ku, Tokyo 113-0033, Japan.}

\begin{abstract}
Nested error regression models are useful tools for analysis of grouped data, especially in the case of small area estimation.
This paper suggests a nested error regression model using uncertain random effects in which the random effect in each area is expressed as a mixture of a normal distribution and a positive mass at $0$.
For estimation of the model parameters and prediction of the random effects, an objective Bayesian inference is proposed by setting non-informative prior distributions on the model parameters.
Under mild sufficient conditions, it is shown that the posterior distribution is proper and the posterior variances are finite, confirming the validity of posterior inference.
To generate samples from the posterior distribution, we provide the Gibbs sampling method with familiar forms for all the full conditional distributions.
This paper also addresses the problem of predicting finite population means, and a sampling-based method is suggested to tackle this issue.
Finally, the proposed model is compared with the conventional nested error regression model through simulation and empirical studies.
\end{abstract}

\begin{keyword}
Bayesian estimator, nested error regression model, posterior propriety, small area estimation, uncertain random effect
\end{keyword}

\end{frontmatter}

\section{Introduction}
Linear mixed models and model-based estimators including the empirical Bayes estimator (EB) or empirical best linear unbiased predictor (EBLUP) have been studied quite extensively in the literature from both theoretical and applied points of view.
Of these, small area estimation is an important application, and methods for small area estimation have received much attention in recent years due to a growing demand for reliable small area estimates.
For good reviews on this topic, see Ghosh and Rao \cite{GR}, Rao and Molina \cite{Rao}, Datta and Ghosh \cite{DM} and Pfeffermann \cite{P}.
The linear mixed models used for small area estimation are categorized into two major types, the Fay-Herriot model suggested by Fay and Herriot \cite{FH} for area-level data, and the nested error regression (NER) models given in Battese, Harter and Fuller \cite{BHF} for unit-level data.
The resulting model-based estimators, such as EB or EBLUP, for small-cluster means or subject-specific values,  provide reliable estimates with higher precision than direct estimates like sample means. 
These stable inferences are owing to random effects, but the misspecification of random effects may increase the risk of prediction.

Concerning this issue, Datta, Hall and Mandal \cite{DHM} recently suggested inference by testing the presence of random effects in general mixed models.
They pointed out that if the random effects can be dispensed, the model parameters and the small area means may be estimated with substantially higher accuracy.
Further, Datta and Mandal \cite{DM} generalized the idea of preliminary testing to the uncertain random effects in the Fay-Herriot model, which is described as 
$$
y_i=\th_i+\ep_i, \ \ \ \th_i=\x_i^{\top}\bbe+u_iv_i, \ \ \ \ i=1,\ldots,m,
$$
where $\ep_i\sim \mathcal{N}(0,D_i)$ for known $D_i$, $v_i\sim \mathcal{N}(0,A)$ and $\Pr(u_i=1)=p=1-\Pr(u_i=0)$.
In Datta and Mandal \cite{DM}, the term $u_iv_i$ is called the ``uncertain random effect" since the density of $u_iv_i$ is expressed as a mixture of $\mathcal{N}(0,A)$ and the one-point distribution on $0$.
The mixture expression of the distribution of random effects can control the extent of random effects and flexible prediction can be achieved.
Actually, the resulting estimator (predictor) of $\th_i$ is expressed as the linear combination of the direct estimator $y_i$ and the regression estimator $\x_i^{\top}\bbeh$.
The weight depends on the squared residuals $(y_i-\x_i^{\top}\bbeh)^2$ while the weight in the resulting estimator from the traditional Fay-Herriot model does not take the residuals into account. 
In Datta and Mandal \cite{DM}, the Bayesian method was implemented for inferences of the small area parameters $\th_i$'s as well as the model parameters by setting the proper prior distributions for $p$ and $A$, namely $p\sim Beta(a_1,a_2)$ and $A\sim IG(a_3,a_4)$ for known (user specified) $a_i, \ i=1,2,3,4$, and the improper uniform prior for $\bbe$, where $Beta(a_1, a_2)$ and $IG(a_3,a_4)$ denote the beta and inverse gamma distributions, respectively.
It was shown that the resulting posterior distributions of all the parameters are proper under some conditions.
However, Datta and Mandal \cite{DM} focused on the Fay-Herriot model, and their method could be restrictive in real applications.
Moreover, they used the proper (informative) prior distribution for both $p$ and $A$, and the result could be affected by the choice of hyperparameters.

In this paper, we treat not only the uncertain random effects in more general small area models like the NER model, but also non-informative prior distributions for model parameters.
The NER model has been used in various applications including small area estimation, biological experiments and econometric analysis.
The NER model is described as 
$$
y_{ij}=\x_{ij}^{\top}\bbe+v_i+\ep_{ij}, \ \ \ j=1,\ldots,n_i, \ \ i=1,\ldots,m,
$$ 
where $\ep_{ij}$ is the sampling error associated with $y_{ij}$ and $v_i$ is a random effect in the $i$th area.
It is usually assumed that $\ep_{ij}$ and $v_i$ are mutually independent and distributed as $\ep_{ij}\sim \mathcal{N}(0,\si^2)$ and $v_i\sim \mathcal{N}(0,\tau^2)$, respectively.
The main purpose of the NER model is to predict (estimate) the quantity of linear combinations of $\bbe$ and $v_i$, namely $\mu_i=\c_i^{\top}\bbe+v_i$ for some known vector $\c_i$.
For a decade, there has been criticism that the assumption of the NER model is not necessarily satisfied in real applications and several extensions of the NER model have been proposed in order to adapt to real data sets.
For example, Jiang and Nguyen \cite{JN}, Kubokawa, Sugasawa, Ghosh and Chaudhuri \cite{KSGC} and Sugasawa and Kubokawa \cite{DM} proposed heteroscedastic nested error regression models in which the variance components $\tau^2$ and $\si^2$ are not constant over the areas.
Also, Ghosh, Sinha and Kim \cite{GSK}, Arima, Datta and Liseo \cite{ADL} and Torabi \cite{Torabi} introduced extended models with measurement errors in covariates.
However, the problem of uncertainty of random effects, to our knowledge, has not been considered so far in the context.

In this article, we suggest the use of the uncertain random effect in the NER model and propose the uncertain nested error regression (UNER) model by adopting the structure 
$$
v_i|u_i\sim \mathcal{N}(0,u_i\tau^2)\quad {\rm with}\quad \Pr(u_i=1)=p.
$$
For the prior distribution of $\tau^2$, the variance of random effects, we use the prior distribution depending on $u_i$'s, which is defined as
$$
\pi(\tau^2|z>a)\propto \tau^{-1}, \ \ \ 
\pi(\tau^2|z\leq a)\propto \pi_{\ast}(\tau^2),
$$
for some $a>0$, where $z=\sum_{i=1}^mu_i$ and $\pi_{\ast}(\tau^2)$ is some proper density, so that the prior distribution of $\tau^2$ is more non-informative than the proper prior such as an inverse gamma distribution as used in Datta and Mandal \cite{DM}.
For the other parameters $\bbe,\si^2$ and $p$, we also assign the non-informative prior as $\pi(\bbe,\si^2,p)\propto p^{-1/2}(1-p)^{-1/2}\si^{-1}$.
Hence, our Bayesian procedure is objective.
We also apply the NER model in the framework of the finite population to predict the true finite population mean based on the partially observed data in each population.

This article is organized as follows. 
In Section \ref{sec:model}, we describe the details of the UNER model and provide the Bayesian estimation method as well as the main theorem regarding the propriety of the posterior distribution and the finiteness of posterior variances.
The prediction problem of finite population means using UNER is also discussed.
In Section \ref{sec:num}, we compare the UNER model with the NER model through simulation and empirical studies.
Concluding remarks are given in Section \ref{sec:conc} and the technical proof is given in the Appendix.

\section{Uncertain Nested Error Regression Models}\label{sec:model}

\subsection{Model settings and Bayes estimator}\label{sec:setting}
We consider the following uncertain nested error regression (UNER) model 
\begin{equation}\label{model}
\begin{split}
&y_{ij}=\x_{ij}^{\top}\bbe+v_i+\ep_{ij}, \ \ \ j=1,\ldots,n_i,\\
&v_i|(u_i=1)\sim \mathcal{N}(0,\tau^2), \ \ \ v_i|(u_i=0)\sim \de_0(v_i), \ \ \ \ i=1,\ldots,m,
\end{split}
\end{equation}
independently for $i$ with $\Pr(u_i=1)=1-\Pr(u_i=0)=p$, where $\x_{ij}$ is a $q$-dimensional vector of covariates, $\bbe$ is a $q$-dimensional vector of regression coefficients, $\de_0(\cdot)$ denotes the Dirac measure on $0$, and $\ep_{ij}$'s are independently and identically distributed as $\mathcal{N}(0,\si^2)$.
The marginal density function of $v_i$ is given by
$$
f(v)=\frac{p}{\sqrt{2\pi}\tau}\exp\Big(-\frac{v^2}{2\tau^2}\Big)+(1-p)I(v=0), 
$$
which is a mixture of the normal distribution $\mathcal{N}(0,\tau^2)$ and the point mass on $0$.
Thus the model parameters are regression coefficients $\bbe$, the variance components $\si^2$ and $\tau^2$, and the mixture ratio $p$.
Let $\y_i=(y_{i1},\ldots,y_{in_i})^{\top}$ be the observed vector in the $i$th area.
Then the variance of $\y_i$ is $\Var(\y_i)=\si^2\I_{n_i}+p\tau^2\J_{n_i}$ for $\J_{n_i}=\1_{n_i}\1_{n_i}^{\top}$.
If the prior probability $p$ of $u_i=1$ is $0$, it follows that $\Var(\y_i)=\si^2\I_{n_i}$, and the observations in the $i$th area are mutually independent.
The parameter which we want to estimate (predict) is $\mu_i=\c_i^{\top}\bbe+v_i$ for a known vector $\c_i$.
The typical choice of $\c_i$ is $\barx_i=n_i^{-1}\sum_{j=1}^{n_i}\x_{ij}$ in which $\mu_i$ corresponds to the mean of the $i$th area.

The posterior distribution of $\mu_i$ given $u_i$ and $\y_i$ is 
$$
\mu_i|u_i,\y_i\sim N\Big(\c_i^{\top}\bbe+\frac{n_i\tau^2I(u_i=1)}{\si^2+n_i\tau^2}(\bary_i-\barx_i^{\top}\bbe),\ \frac{ I(u_i=1)\si^2\tau^2}{\si^2+n_i\tau^2}\Big),
$$
where $\bary_i=n_i^{-1}\sum_{j=1}^{n_i}y_{ij}$, the sample mean of $y_{ij}$ in the $i$th area.
Thus the posterior distribution of $\mu_i$ given $\y_i$ is a mixture of the normal distribution and one point mass on $\c_i^{\top}\bbe$.
The resulting Bayes estimator $\mut_i$ of $\mu_i$ is 
\begin{align*}
\mut_i=\E[\mu_i|\y_i]&=\pt_i\Big\{\c_i^{\top}\bbe+\frac{n_i\tau^2}{\si^2+n_i\tau^2}(\bary_i-\barx_i^{\top}\bbe)\Big\}+(1-\pt_i)\c_i^{\top}\bbe\\
&=\c_i^{\top}\bbe+\frac{n_i\tau^2\pt_i}{\si^2+n_i\tau^2}(\bary_i-\barx_i^{\top}\bbe),
\end{align*}
where $\pt_i$ is the posterior probability of $u_i=1$ given by
\begin{equation}\label{prob}
\begin{split}
\pt_i&=\Pr(u_i=1|\y_i)\\
&=p\Big\{p+(1-p)\sqrt{\frac{\si^2+n_i\tau^2}{\si^2}}\exp\Big(-\frac{n_i^2\tau^2}{2\si^2(\si^2+n_i\tau^2)}(\bary_i-\barx_i^{\top}\bbe)^2\Big)\Big\}^{-1}.
\end{split}
\end{equation}
We note that $\pt_i$ increases in $p$ and $(\bary_i-\barx_i^{\top}\bbe)^2$.
Thus, if $\x_{ij}$ is a good covariate to explain $y_{ij}$ in the $i$th area, the squared residual $(\bary_i-\barx_i^{\top}\bbe)^2$ is expected to be small, and the posterior probability $\pt_i$ is small as well.
The posterior probability $\pt_i$ is $1$ when $p=1$ and $\pt_i$ converges to $1$ as $(\bary_i-\barx_i^{\top}\bbe)^2$ goes to infinity.

Moreover, the posterior variance of $\mu_i$ is expressed as
\begin{equation}\label{pos.var}
\begin{split}
V_i(\y_i)&\equiv\Var(\mu_i|\y_i)=\Var(v_i|\y_i)\\
&=\frac{n_i^2\tau^4}{(\si^2+n_i\tau^2)^2}(\bary_i-\barx_i^{\top}\bbe)^2\pt_i(1-\pt_i)+\frac{\si^2\tau^2\pt_i}{\si^2+n_i\tau^2}.
\end{split}
\end{equation}
It is interesting to point out that the posterior variance of $\mu_i$, in this case, depends on observation $\y_i$ through the squared residual $(\bary_i-\barx_i^{\top}\bbe)^2$ and the posterior probability $\pt_i$, while the posterior variance of the random effect in the usual nested error regression model is given by $\si^2\tau^2(\si^2+n_i\tau^2)^{-1}$, which does not depend on observation $\y_i$.
This means that the uncertain random effect enables us to take the distance between sample mean $\bary_i$ and synthetic estimator $\barx_i^{\top}\bbe$ into the posterior variability of the interesting parameter $\mu_i$.

\subsection{Bayesian implementation and posterior distribution}\label{sec:post}
Since the marginal likelihood function of the model parameters $\bbe,\si^2,\tau^2$ and $p$ is rather complex, we consider objective Bayesian inference for the model parameters as well as the random effect $v_i$.
To this end, we rewrite the model (\ref{model}) as 
\begin{equation}\label{unerm}
\begin{split}
&y_{ij}|v_i,\bbe,\si^2\sim \mathcal{N}(\x_{ij}^{\top}\bbe+v_i,\si^2), \ \ \ j=1,\ldots,n_i, \ \ i=1,\ldots,m\\
&v_i|u_i,\tau^2\sim \mathcal{N}(0,u_i\tau^2), \ \ \ u_i|p\sim Ber(p), \ \ \ i=1,\ldots,m
\end{split}
\end{equation}
independently for $i$, where $Ber(p)$ denotes the Bernoulli distribution.
For implementation of full Bayesian inference, we need to set prior distributions on the model parameters.
To keep objectivity of inferences, we use the uniform prior distribution on $\bbe$ and the Jeffreys prior distributions on $\si^2$ and $p$.
On the other hand, the prior distribution of $\tau^2$ should depend on $z=\sum_{i=1}^mu_i$, since $\tau^2$ cannot be identified for a small value of $z$.
Thus, for the model parameters, we use the prior distributions
\begin{equation}\label{prior}
\pi(\bbe,\si^2,p)=p^{-1/2}(1-p)^{-1/2}\si^{-1}, \ \ \pi(\tau^2|z)\propto 
\begin{cases}
\tau^{-1} & (z>a)\\
\pi_{\ast}(\tau^2) & (z\leq a)
\end{cases}
\end{equation}
where $\pi_{\ast}(\tau^2)=(\tau^2)^{-b_1-1}\exp(-b_2/\tau^2)$ for known constants $b_1>3$ and $b_2>0$.
The value of $a$ is chosen by the user, and this point will be discussed later.
It is noted that the prior distribution on $p$ is proper, but the priors on $\bbe,\si^2$ and $\tau^2$ are improper, so that the posterior propriety is not always guaranteed.
In Theorem \ref{thm:pos}, we show that the posterior distribution for the model parameters is proper under mild conditions.

We now describe the posterior distribution and investigate its properties. 
The set of all observed data is denoted by $D=\{\y_i,\X_{i}\}_{i=1,\ldots,m}$ for $\X_i=(\x_{i1},\ldots,\x_{in_i})$.
From the model (\ref{unerm}) with prior setup (\ref{prior}), the posterior density of parameters $(\v,\u,\bbe,\si^2,\tau^2,p)$ for $\v=(v_1,\ldots,v_m)^{\top}$ and $\u=(u_1,\ldots,u_m)^{\top}$ is given by
\begin{equation}\label{pos}
\begin{split}
\pi&(\v,\u,\bbe,\si^2,\tau^2,p|D)\\
\propto& (\si^2)^{-(N+1)/2}(\tau^2)^{-(z+I(z>a))/2-(b_1+1)I(z\leq a)}p^{z-1/2}(1-p)^{m-z-1/2}\\
&\times \prod_{i=1}^m\Big[ \exp\Big(-\frac{\sum_{j=1}^{n_i}(y_{ij}-\x_{ij}^{\top}\bbe-v_i)^2}{2\si^2}-\frac{u_iv_i^2}{2\tau^2}\Big)\de_0(v_i)^{1-u_i}\Big]
\\
&\times \exp\Big\{-\frac{b_{2}}{\tau^{2}}I(z\leq a)\Big\}.
\end{split}
\end{equation}

Now, we state our main result about the posterior propriety and the existence of posterior variances .

\begin{thm}\label{thm:pos} 
 {\rm (a)}\ \ The marginal posterior density $\pi(\bbe,\si^2,\tau^2,p|D)$ is proper if $N>q+2$ and $m> a\geq 1$\\
{\rm (b)}\ \ The model parameters $\bbe,\si^2,\tau^2$ and $p$ have finite posterior variances if $N>q+6$ and $m>a\geq 5$.
\end{thm}

Remember that $q$ is the dimension of the vector of regression coefficients $\bbe$, and $a$ is the tuning parameter of the prior for $\tau^2$.
Part (a) in Theorem \ref{thm:pos} says that the marginal posterior densities of the small area means are proper and part (b) provides a sufficient condition for obtaining finite measures of uncertainty for the model parameters. 
We note that the conditions in Theorem \ref{thm:pos} are similar to the conditions given in Arima, et al. \cite{ADL} and Datta and Mandal \cite{DM}.
The proof of Theorem \ref{thm:pos} is presented in the Appendix.

Since the posterior distribution in (\ref{pos}) cannot be obtained in a closed form, we rely on the Markov chain Monte Carlo technique, in particular the Gibbs sampler, in order to draw samples from the posterior distribution.
This requires generating samples from the full conditional distributions for each of $(\v,\u,\bbe,\si^2,\tau^2,p)$ given the remaining parameters and the data $D$. 
Fortunately, the full conditional distributions are described as familiar distributions allowing us to easily implement the Gibbs sampling.
The full conditional distributions are given by
\begin{equation}\label{full}
\begin{split}
&v_i|u_i,\bbe,\si^2,\tau^2,D\sim N\Big(\frac{n_i\tau^2I(u_i=1)}{\si^2+n_i\tau^2}(\bary_i-\barx_i^{\top}\bbe), \frac{\si^2\tau^2I(u_i=1)}{\si^2+n_i\tau^2}\Big),  \ \ i=1,\ldots,m,\\
&u_i|\bbe,\si^2,\tau^2,p,D\sim Ber(\pt_i), \ \ \ i=1,\ldots,m, \ \ \ \ \ 
p|\u,D\sim Beta\Big(z+\frac12,m-z+\frac12\Big), \\
&\bbe|\u,\si^2,\tau^2,D\sim N_p((\X^{\top}\bSi_u^{-1}\X)^{-1}\X^{\top}\bSi_u^{-1}\y,(\X^{\top}\bSi_u^{-1}\X)^{-1}), \\
&\tau^2|\u,\v,D\sim IG\Big(\frac12 (z-I(z>a))+b_{1}I(z\leq a),\frac12\sum_{i=1}^mu_iv_i^2+b_{2}I(z\leq a)\Big),\\
&\si^2|\v,\bbe,D\sim IG\Big(\frac12(N-1),\frac12(\y-\X\bbe-\Z\v)^{\top}(\y-\X\bbe-\Z\v)\Big), \ \ \ 
\end{split}\end{equation}
where $z=\sum_{i=1}^mu_i$, $\bSi_u=\diag(\bSi_{1u},\ldots,\bSi_{mu})$ with $\bSi_{iu}=\si^2 \I_{n_i}+u_i\tau^2\1_{n_i}\1_{n_i}^{\top}$, $\y=(\y_1^{\top},\ldots,\y_m^{\top})^{\top}$, $\X=(\X_1,\ldots,\X_m)$, and $\pt_i$ is given in (\ref{prob}).
Using these expressions of full conditional distributions, we can easily draw posterior samples of all the variances and parameters to make inferences, such as point estimation, prediction intervals and standard errors, for $\mu_i=\c_i^{\top}\bbe+v_i$.

In the closing of this section, we discuss the choices of $a, b_1$ and $b_2$ in the posterior distribution of $\tau^2$.
We remember that the prior distribution of $\tau^2$ is non-informative and improper when $z>a$ and informative and proper when $z\leq a$.
Taking it into account, we should select a value of $a$ as small as possible.
Hence, it follows from Theorem \ref{thm:pos} that $a=5$ is the most reasonable choice.
On the other hand, as discussed in Datta and Mandal \cite{DM}, a reasonable choice is $b_1=V+2$ and $b_2=V(V+1)$ such that $\E[\tau^2|z\leq a]=V$ and $\Var(\tau^2|z\leq a)=V^2$, where $V$ is the estimated sampling variance given by
$$
V=\frac1{N-m-q}\sum_{i=1}^m\sum_{j=1}^{n_i}\{y_{ij}-\bary_i-(\x_{ij}-\barx)^{\top}\bbeh_{\rm OLS}\}^2.
$$
Here, $\bbeh_{\rm OLS}$ is the ordinary least squared estimator of $\bbe$.
It should be noted that $V$ satisfies $\E[V]=\si^2$.

\subsection{Prediction in finite populations}\label{sec:fp}
Here, we consider the problem of predicting the means in finite populations.
Assume that there exist $m$ finite populations and the $i$th population consists of $N_i$ pairs of data $(Y_{ij},\x_{ij})$, $j=1,\ldots,N_i$.
It is supposed that $n_i (<N_i)$ observations are sampled from the $i$th population.
What we want to predict is the mean of the $i$th finite population $\barY_i=N_i^{-1}\sum_{j=1}^{N_i}Y_{ij}$.
Assume also that the mean vector of covariates $\barX_i=N_i^{-1}\sum_{j=1}^{N_i}\x_{ij}$ is available, which is often encountered in real application (Battese, et al. \cite{BHF}).
Let $s_i$ and $r_i$ be collections of indices of sampled and non-sampled observations in the $i$th area, respectively, so that $s_i$ and $r_i$ satisfy $s_i\cap r_i=\phi$ and $s_i\cup r_i=\{1,\ldots N_i\}$.
Without loss of generality, we assume that $s_i=\{1,\ldots,n_i\}$ and $r_i=\{n_i+1,\ldots,N_i\}$.
The Bayes estimator of $\barY_i$ under quadratic loss is given by 
$$
\E[\barY_i|\y_i]=\frac{1}{N_i}\Big\{n_i\bary_{i(s)}+(N_i-n_i)\E[\barY_{i(r)}|\y_i]\Big\},
$$
where
$$
\bary_{i(s)}=n_i^{-1}\sum_{j\in s_i}y_{ij}, \ \ \  \barY_{i(r)}=(N_i-n_i)^{-1}\sum_{j\in r_i}Y_{ij}.
$$
For evaluating the conditional mean $\E[\barY_{i(r)}|\y_i]$, we assume that $Y_{ij}$ is expressed as
$$
Y_{ij}=\x_{ij}^{\top}\bbe+v_i+\ep_{ij}, \ \ j\in r_i,
$$
that is, the non-sampled observations have the same data generating structure as the sampled ones.
Then the unobserved mean $\barY_{i(r)}$ is expressed as
$$
\barY_{i(r)}=\barx_{i(r)}^{\top}\bbe+v_i+\barep_{i(r)},
$$
where $\barep_{i(r)}=(N_i-n_i)^{-1}\sum_{j\in r_i}\ep_{ij}$.
Thus the conditional distribution of $\barY_{i(r)}$ given $\y_i$ and $u_i$ is 
\begin{equation}\label{fp-cond}
\barY_{i(r)}|\y_i,u_i\sim N\Big(\barx_{i(r)}^{\top}\bbe+\frac{I(u_i=1)n_i\tau^2}{\si^2+n_i\tau^2}(\bary_i-\barx_i^{\top}\bbe), \ \frac{I(u_i=1)\si^2\tau^2}{\si^2+n_i\tau^2}+\frac{\si^2}{N_i-n_i}\Big),
\end{equation}
which yields the predictive density of $\barY_{i(r)}$ given by
\begin{align*}
\barY_{i(r)}|\y_i
\sim& \pt_iN\Big(\barx_{i(r)}^{\top}\bbe+\frac{n_i\tau^2}{\si^2+n_i\tau^2}(\bary_i-\barx_i^{\top}\bbe), \ \frac{\si^2\tau^2}{\si^2+n_i\tau^2}+\frac{\si^2}{N_i-n_i}\Big)\\
& \ \ \ +(1-\pt_i)N\Big(\barx_{i(r)}^{\top}\bbe,\frac{\si^2}{N_i-n_i}\Big),
\end{align*}
where $\pt_i$ is the posterior probability of $u_i=1$ given in (\ref{prob}).
Thus the conditional distribution of the non-sampled data is a mixture of the two normal distributions of the predictive density, with and without random effect.
Moreover, the conditional variance $\barY_{i(r)}$ given $\y_i$ is calculated as $V_i(\y_i)+(N_i-n_i)^{-1}\si^2$, where $V_i(\y_i)$ is the posterior variance of $v_i$ given in (\ref{pos.var}).
It is noted that, when the true mean vector of the explanatory variables $\barX_i$ is available in each area, the value of $\barx_{i(r)}$ is easily obtained by 
$$
\barx_{i(r)}=(N_i-n_i)^{-1}(N_i\barX_i-n_i\barx_i).
$$
To implement the prediction in the finite population model, we regard $\barY_{i(r)}$ as latent variables and add the sampling step from (\ref{fp-cond}) to the Gibbs sampling given in (\ref{full}).

\section{Numerical Studies}\label{sec:num}

\subsection{Model based simulations}
In this simulation study, we compared the UNER model with the conventional NER model in terms of the quality of the estimates.
In applying the NER model, we used the Jeffreys prior on $(\bbe,\tau^2,\si^2)$, namely $\pi(\bbe,\tau^2,\si^2)=\tau^{-1}\si^{-1}$, where it is well-known that the resulting posterior distribution is proper (Berger \cite{Berger}).
The full conditional posterior distributions are given by
\begin{equation}\label{posNER}
\begin{split}
&v_i|\bbe,\si^2,\tau^2,D\sim N\Big(\frac{n_i\tau^2}{\si^2+n_i\tau^2}(\bary_i-\barx_i^{\top}\bbe), \frac{\si^2\tau^2}{\si^2+n_i\tau^2}\Big),\ \ \ \ i=1,\ldots,m\\
&\bbe|\tau^2\si^2,D\sim N_p((\X^{\top}\bSi^{-1}\X)^{-1}\X^{\top}\bSi^{-1}\y,(\X^{\top}\bSi^{-1}\X)^{-1}), \\
&\tau^2|\v,D\sim IG\Big(\frac12(m-1),\frac12\sum_{i=1}^mv_i^2\Big),\\
&\si^2|\v,\bbe,D\sim IG\Big(\frac12(N-1),\frac12(\y-\X\bbe-\Z\v)^{\top}(\y-\X\bbe-\Z\v)\Big),
\end{split}\end{equation}
where $\bSi=\diag(\bSi_1,\ldots,\bSi_m)$ with $\bSi_i=\si^2\I_{n_i}+\tau^2\1_{n_i}\1_{n_i}^{\top}$.
We considered the following data generating process:
$$
y_{ij}=\beta_0+\beta_1x_{ij}+v_i+\ep_{ij}, \ \ \ \ j=1,\ldots,n, \ \ \ \ i=1,\ldots,m,
$$
where $\ep_{ij}\sim \mathcal{N}(0,1)$, $\beta_0=1$, $\beta_1=0.5$, and $x_{ij}$'s were generated from the uniform distribution on $(1,2)$ and fixed through simulation runs.
The four combinations of $(n,m)$ were considered as $(n,m)=(5,20)$, $(5,40)$, $(10,20)$, $(10,40)$.
For the true distributions of $v_i$, we considered the following four scenarios for each choice of $(n,m)$.
\begin{align*}
&{\rm S1}: \  v_i\sim \mathcal{N}(0,(0.7)^2), \ \ \ \  
{\rm S2}: \ v_i\sim 0.3\de_0(v_i)+0.7\mathcal{N}(0,(0.7)^2), \\
&{\rm S3}: \ v_i\sim 0.3\de_0(v_i)+0.7\mathcal{L}(0,(0.7)^2), \ \ \ \ 
{\rm S4}: \ v_i\sim 0.3\de_0(v_i)+0.7t_6(0,(0.7)^2),
\end{align*}
where $t_6(a,b)$ and $\mathcal{L}(a,b)$ denote the scaled $t$-distribution with $6$ degrees of freedom with mean $a$ and variance $b$ and the scaled Laplace distribution with mean $a$ and variance $b$, respectively.
Hence, UNER is misspecified in scenarios S3 and S4, and overspecified in scenario S1.

Based on $R=1,000$ simulation runs, we computed the mean squared errors (MSE), absolute bias of $\muh_i$, and empirical coverage probability of the $95\%$ credible interval of $\mu_i$, which are respectively defined as 
\begin{align*}
&{\rm MSE}=\frac{1}{mR}\sum_{r=1}^R\sum_{i=1}^m(\muh_i^{(r)}-\mu_i^{(r)})^2, \ \ \ \ 
{\rm Bias}=\frac{1}{mR}\sum_{r=1}^R\sum_{i=1}^m|\muh_i^{(r)}-\mu_i^{(r)}|\\
&{\rm CP}=\frac{1}{mR}\sum_{r=1}^R\sum_{i=1}^mI(\mu_i^{(r)}\in {\rm CI}_i^{(r)})\times 100,
\end{align*}
where $\muh_i^{(r)}$, $\mu_i^{(r)}$ and ${\rm CI}_i^{(r)}$ are the posterior mean, the true value, and the $95\%$ credible intervals, respectively, of $\mu_i$ in the $r$th simulation runs.
In each iteration of the simulation run, we used $5,000$ posterior samples after $1,000$ initial iterations for both UNER and NER. 
The results are given in Table \ref{tab:sim}.
In scenario S1, both the MSE and absolute bias of UNER are larger than those of NER since UNER is overspecified.
However, as the number of $n$ and $m$ get large, the difference of these values gets small.
For the other scenarios, we can observe that UNER clearly performs better than NER in terms of MSE and absolute bias, and the differences get larger as $n$ and $m$ get larger.
Finally, it is observed that the coverage probability of credible intervals are similar in UNER and NER.
Hence, we can conclude that UNER is expected to be a useful tool when $m$ and $n$ are moderate or large.

\begin{table}
\caption{Simulated MSE, Bias and Coverage Probabilities (CP) of UNER and NER in Different Scenarios. }
\begin{center}
$
{\renewcommand\arraystretch{1.1}\small
\begin{array}{
c@{\hspace{3mm}}c@{\hspace{3mm}}c@{\hspace{3mm}}
c@{\hspace{3mm}}c@{\hspace{3mm}}c@{\hspace{3mm}}
c@{\hspace{3mm}}c@{\hspace{3mm}}c@{\hspace{3mm}}
c@{\hspace{3mm}}c@{\hspace{3mm}}c@{\hspace{3mm}}
}
\hline
&&& {\rm UNER }&&&& {\rm NER }&\\
(n,m) & {\rm Scenario} &{\rm MSE} & {\rm Bias} &{\rm CP} &&{\rm MSE} & {\rm Bias} &{\rm CP}\\
\hline
(3,25) &{\rm S1} & 0.278 & 0.419 & 92.3 && 0.265 & 0.408 & 92.3\\
&{\rm S2} &  0.165& 0.308 & 93.6 &&  0.176 & 0.320 & 93.4\\
&{\rm S3} &0.156 & 0.293 & 93.3 && 0.166 & 0.309 & 93.2\\
&{\rm S4} &0.163 & 0.301 & 93.9 && 0.172 & 0.313 & 93.8\\
\hline
(3,50) &{\rm S1} & 0.248 & 0.396 & 93.2 && 0.242 & 0.388 & 93.2\\
&{\rm S2} & 0.126 & 0.252 & 94.3 && 0.136 & 0.267 & 94.3\\
&{\rm S3} & 0.128 & 0.245 & 93.6 && 0.140 & 0.261 & 93.6\\
&{\rm S4} & 0.130 & 0.258 & 94.6 && 0.140 & 0.272 & 94.3\\
\hline
(6,25) &{\rm S1} & 0.160 & 0.319 & 93.7 && 0.154 & 0.313 & 93.7\\
&{\rm S2} &  0.088 & 0.215 & 94.1 && 0.098 & 0.235 & 94.1\\
&{\rm S3} & 0.088 & 0.217 & 93.7 && 0.103 & 0.239 & 93.7\\
&{\rm S4} & 0.094 & 0.221 & 93.8 && 0.104 & 0.240 & 93.8\\
\hline
(6,50) &{\rm S1} & 0.144 & 0.302 & 94.3 && 0.141 & 0.299 & 94.3\\
&{\rm S2} & 0.076 & 0.206 & 94.5 && 0.095 & 0.229 & 94.5\\
&{\rm S3} & 0.071 & 0.180 & 94.3 && 0.091 & 0.216 & 94.3\\
&{\rm S4} & 0.077 & 0.191 & 95.1 && 0.088 & 0.216 & 95.1\\
\hline
\end{array}
}
$
\end{center}
\label{tab:sim}
\end{table}

\subsection{Application to PLP data in Japan}
This example, primarily for illustration, used the UNER model (\ref{model}) and data from the posted land price data along the Keikyu train line in 2001.
This train line connects the suburbs in Kanagawa prefecture to the Tokyo metropolitan area.
Those who live in the suburbs in Kanagawa prefecture take this line to work in Tokyo every weekday.
Thus, it is expected that the land price depends on the distance from Tokyo. 
The posted land price data are available for 52 stations on the Keikyu train line, and we consider each station as a small area, namely, $m=52$.
For the $i$th station, data of $n_i$ land spots are available, where $n_i$ varies around $4$ and ranges from $1$ to $11$.

For $j=1, \ldots, n_i$, let $y_{ij}$ denote the log-transformed value of the posted land price (Yen) per for square meter of the $j$th spot, $T_{i}$ is the time it takes from the nearby station $i$ to Tokyo station around 8:30 in the morning, $D_{ij}$ is the geographical distance from spot $j$ to station $i$ and $FAR_{ij}$ denotes the floor-area ratio, or ratio of the building volume to the area at spot $j$. 
These values of $T_i, D_{ij}$ and $FAR_{ij}$ are also transformed  by the logarithmic function. 
We applied the UNER model (\ref{model}) described as
\begin{equation}\label{exm1}
\begin{split}
&y_{ij}=\be_0+FAR_{ij}\be_1+T_{i}\be_2+D_{ij}\be_3+v_i+\ep_{ij},\\
&v_i|(u_i=1)\sim \mathcal{N}(0,\tau^2), \ \ \ v_i|(u_i=0)\sim \delta_{0}(v_i), 
\end{split}
\end{equation}
where $\ep_{ij}$'s are independent and identically distributed as $\mathcal{N}(0,\si^2)$.
For comparison, we also applied the conventional NER model to this data set.

\begin{figure}[!thb]
\centering
\includegraphics[width=6cm,clip]{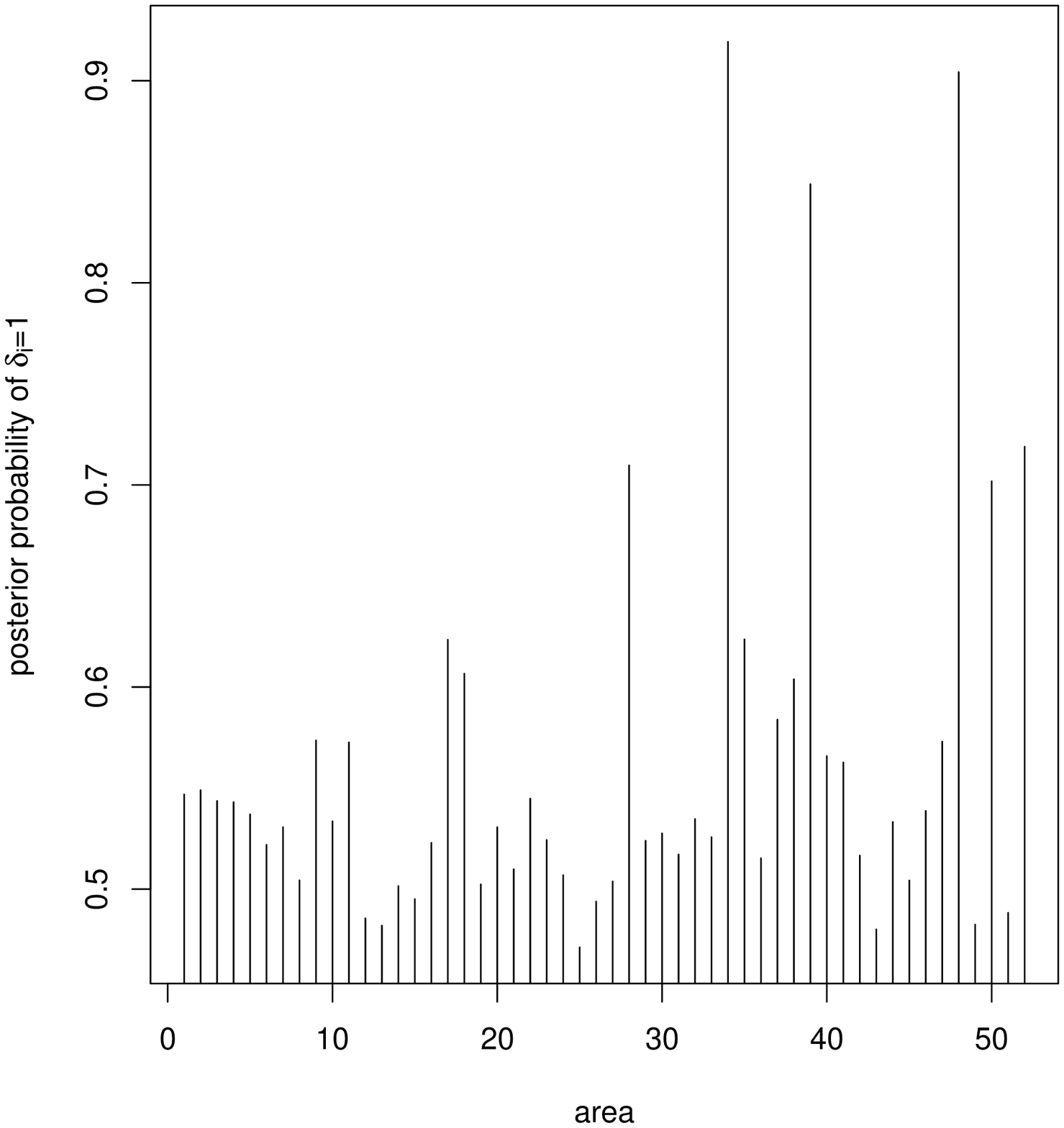}
\includegraphics[width=6cm,clip]{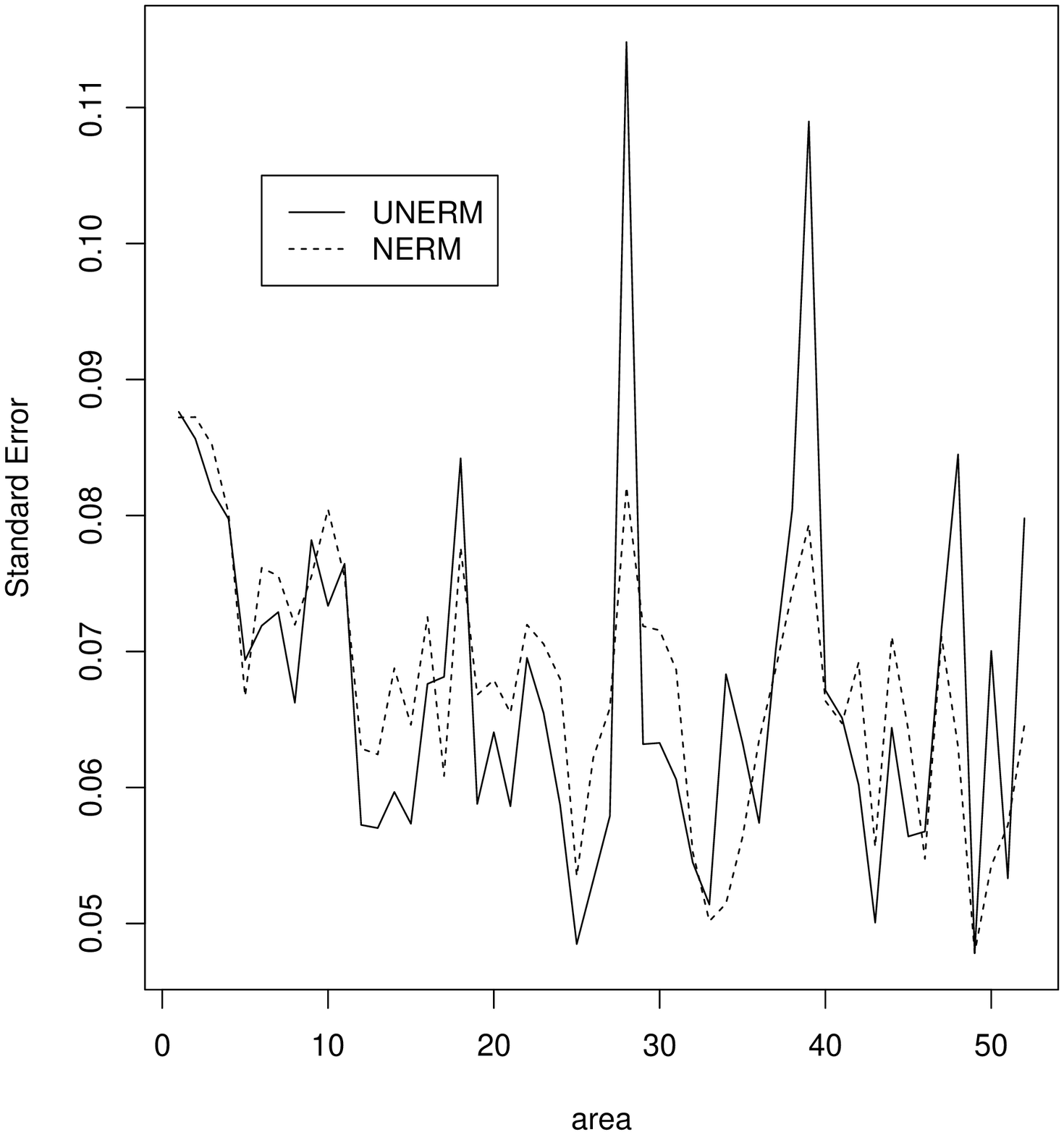}
\caption{Posterior Probability of $u_i=1$ (Left) and Standard Errors of $\mu_i$ (Right) in Each Area.
\label{fig:PLP}
}
\end{figure}

In applying the UNER model, we used the prior distribution with $a=5$ and $b_1=V+2, b_2=V(V+1)$ for $V=0.031$ as discussed in the end of Section \ref{sec:post}.
In both models, we generated $100,000$ posterior samples after $10,000$ iterations of Gibbs sampling given in (\ref{full}) and (\ref{posNER}), respectively, and obtained the posterior means as well as the $95\%$ credible intervals of the model parameters, which are given in Table \ref{tab:PLP}.
Moreover, based on the posterior samples, we computed the Deviance Information Criterion (DIC) suggested in Spiegelhalter, Best, Carlin and van der Linde \cite{DIC}, which is defined as ${\rm DIC}=2\overline{D(\bphi)}-D(\overline{\bphi})$, where $\bphi$ is a vector of the unknown model parameters, $D(\bphi)$ is $(-2)$ times the log-marginal likelihood function, and $\overline{D(\bphi)}$ and $\overline{\bphi}$ denote the posterior means of $D(\bphi)$ and $\bphi$, respectively.
Note that $\bphi=\{\bbe,\tau^2,\si^2,p\}$ in the UNER model, and $\bphi=\{\bbe,\tau^2,\si^2\}$ in the NER model, which are given in Table \ref{tab:PLP} as well.
It is revealed from Table \ref{tab:PLP} that the posterior estimates and credible intervals of regression coefficients $\beta_1,\ldots,\beta_4$ are similar between UNER and NER, and in both models, all the credible intervals of regression coefficients are bounded away from $0$.
On the other hand, the results of variance components $\tau^2$ and $\si^2$ are different because of the effect of the parameter $p$.
In terms of DIC values, the UNER model seems more preferable than the conventional NER model.
To see the effects of $u_i$, we calculated the posterior probabilities $\pt_i$'s which are illustrated in the left panel in Figure \ref{fig:PLP}.
It is revealed that the $\pt_i$'s change dramatically from area to area, and the $\pt_i$'s in most areas are around $0.5$ which comes from the posterior mean of $p=0.54$ as shown in Table \ref{tab:PLP}.

We next considered estimating the land price of a spot with a floor-area ratio of 100$\%$ and a distance of 1000m from the station $i$, namely
$$
\mu_i=\be_0+FAR_0\be_1+T_{i}\be_2+D_0\be_3 +v_i ,
$$
for $FAR_0=\log(100)$ and $D_0=\log(1000)$.
Based on the posterior samples, we calculated the point estimates $\muh_i$ and the posterior standard errors.
The results are given in the right panel of Figure \ref{fig:PLP}, noting that the mean of the posterior standard errors for all areas in UNER and NER are $6.5\times 10^{-2}$ and $6.8\times 10^{-2}$, respectively.
We also computed the length of the prediction intervals of $\mu_i$, and found that the results are similar to standard errors.
It is revealed from Figure \ref{fig:PLP} that UNER provides better estimates than NER in terms of posterior standard errors in most areas.
In some areas, the posterior standard errors of UNER are larger than those of NER when correspondingly the posterior probability $\pt_i$ is larger than $0.7$ as shown in the left panel of Figure \ref{fig:PLP}.
Thus the uncertain random effects may increase the variability of predictors compared to the conventional random effects in the areas where the existence of random effect is strongly supported.
This phenomenon was pointed out in Datta and Mandal \cite{DM} in the Fay-Herriot model.
However, taking the DIC values into account as well, the UNER model works well in this application.

\begin{table}
\caption{Posterior Means and Credible Intervals of the Model Parameters, and DIC.}
\begin{center}
$
{\renewcommand\arraystretch{1.1}\small
\begin{array}{
c@{\hspace{3mm}}c@{\hspace{3mm}}c@{\hspace{3mm}}
c@{\hspace{3mm}}c@{\hspace{3mm}}c@{\hspace{3mm}}
c@{\hspace{3mm}}c@{\hspace{3mm}}c@{\hspace{3mm}}
c@{\hspace{3mm}}c@{\hspace{3mm}}c@{\hspace{3mm}}
}
\hline
  && \text{$\beta_0$} & \text{$\beta_1$} & \text{$\beta_2$} & \text{$\beta_3$}& \text{$\si^2$}   &  \text{$\tau^2$}  & \text{$p$} &&{\rm  DIC   }  \\
\hline
& {\rm 95\%CI\ (upper)}& 15.16 & 0.24 & -0.53 & -0.051 & 0.041& 0.071& 0.99 \\
{\rm UNER}& {\rm mean} &14.55 & 0.17& -0.61  & -0.091    &0.033  &0.017  &0.54 && 512.6\\
& {\rm 95\%CI\ (lower)}& 13.88 & 0.11 & -0.69 & -0.13 1& 0.026 & 0.002 & 0.05\\
\hline
& {\rm 95\%CI\ (upper)}& 15.17 & 0.24 & -0.53 & -0.050 & 0.20 & 0.117 & -\\
{\rm NER}  & {\rm mean} & 14.52 & 0.17 & -0.61 & -0.089 & 0.18 & 0.075 & - && 703.1\\
& {\rm 95\%CI\ (lower)}& 13.88 & 0.10 & -0.69 & -0.132 & 0.16 & 0.031 & -\\
\hline
\end{array}
}
$
\end{center}
\label{tab:PLP}
\end{table}

\subsection{Design based simulation}
\begin{figure}[!thb]
\centering
\includegraphics[width=6cm,clip]{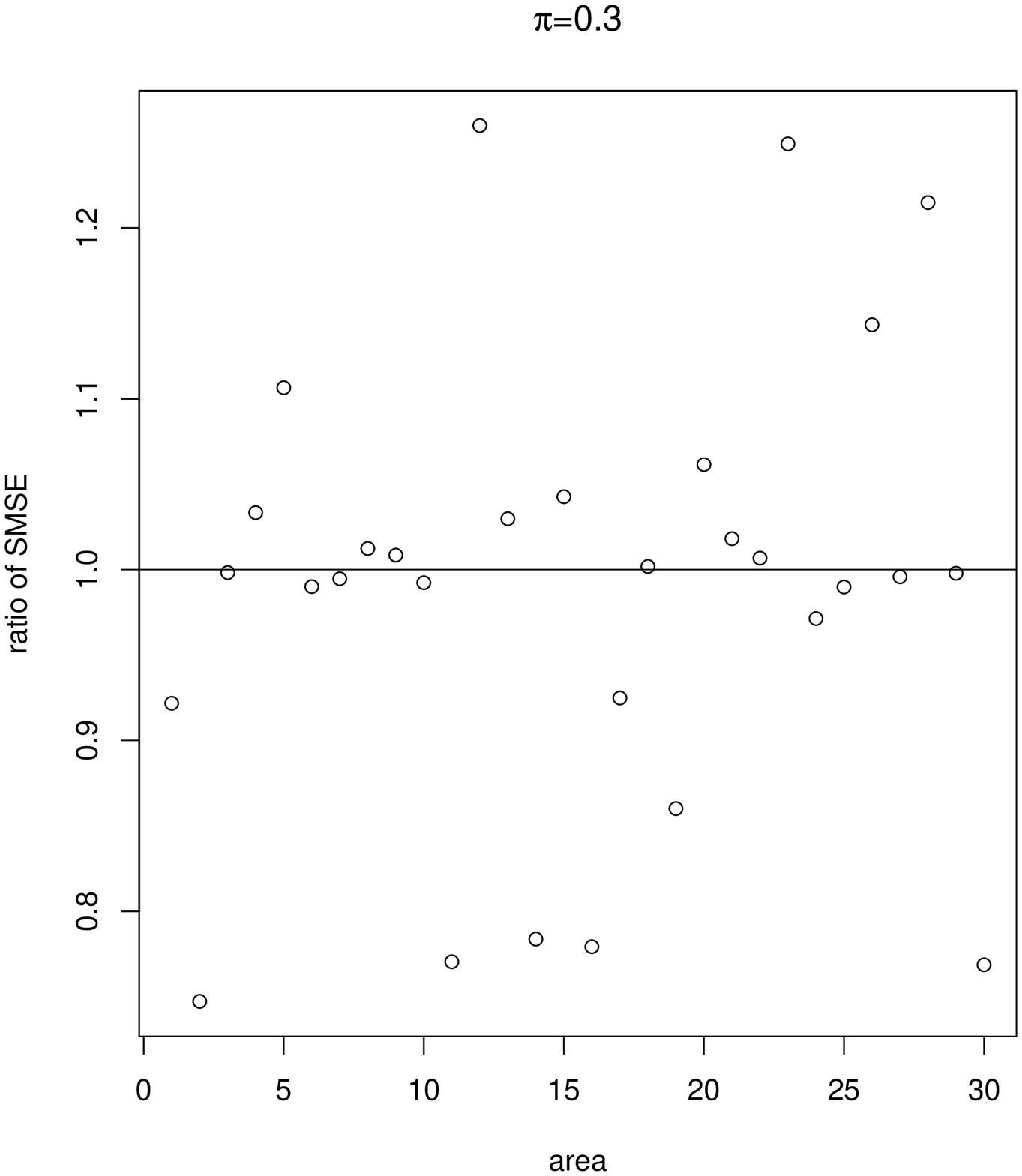}
\includegraphics[width=6cm,clip]{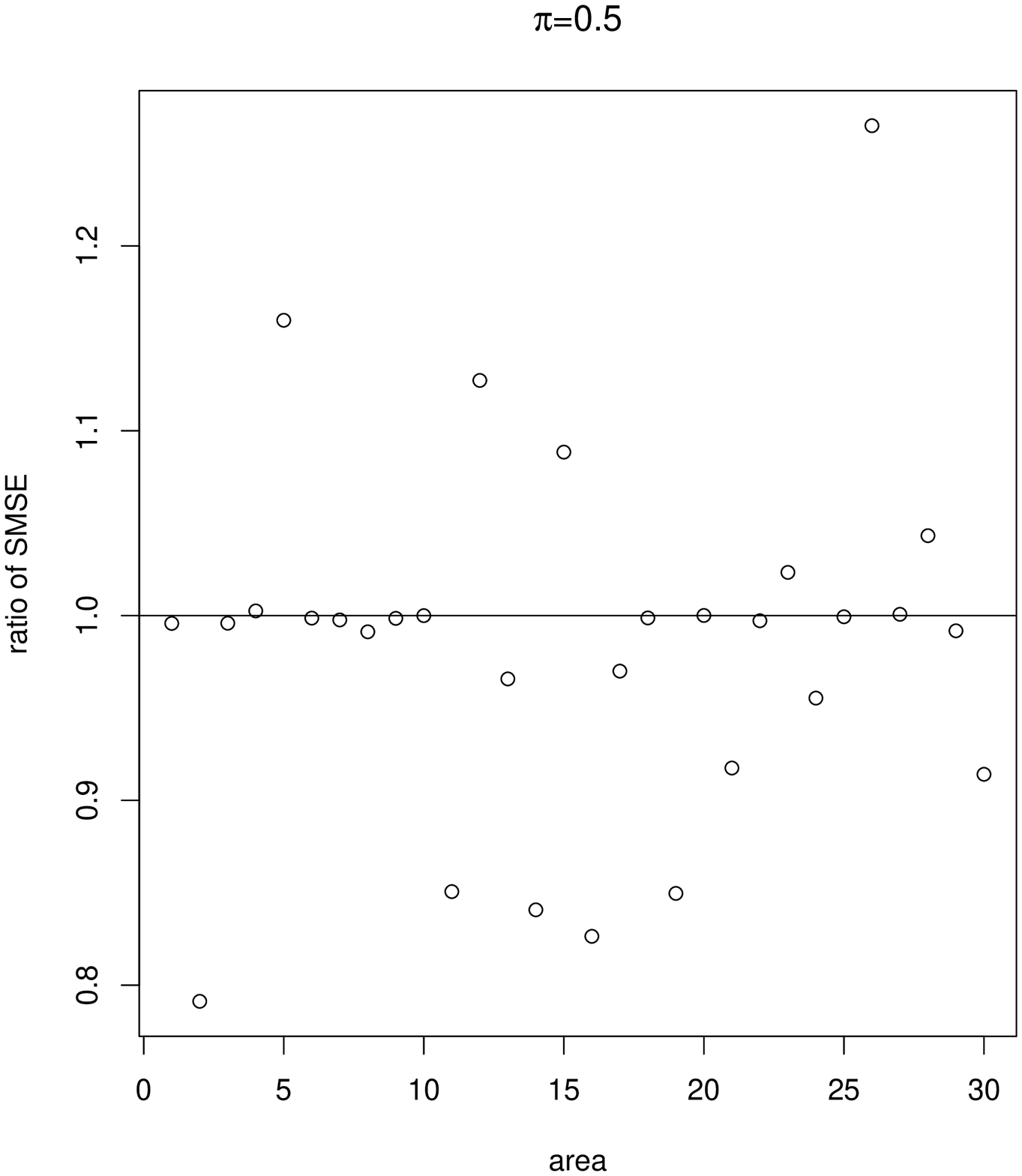}\\
\includegraphics[width=6cm,clip]{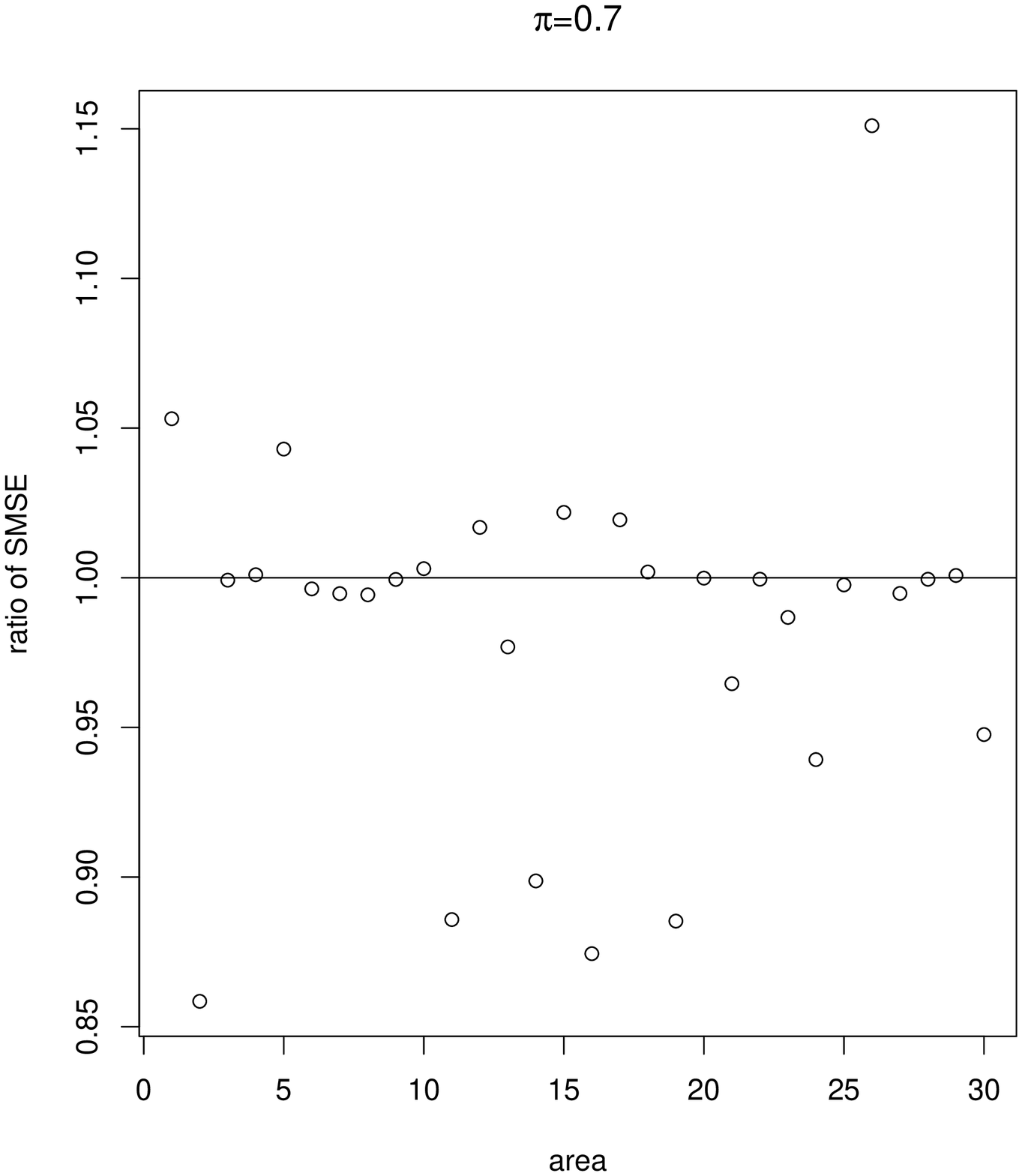}
\includegraphics[width=6cm,clip]{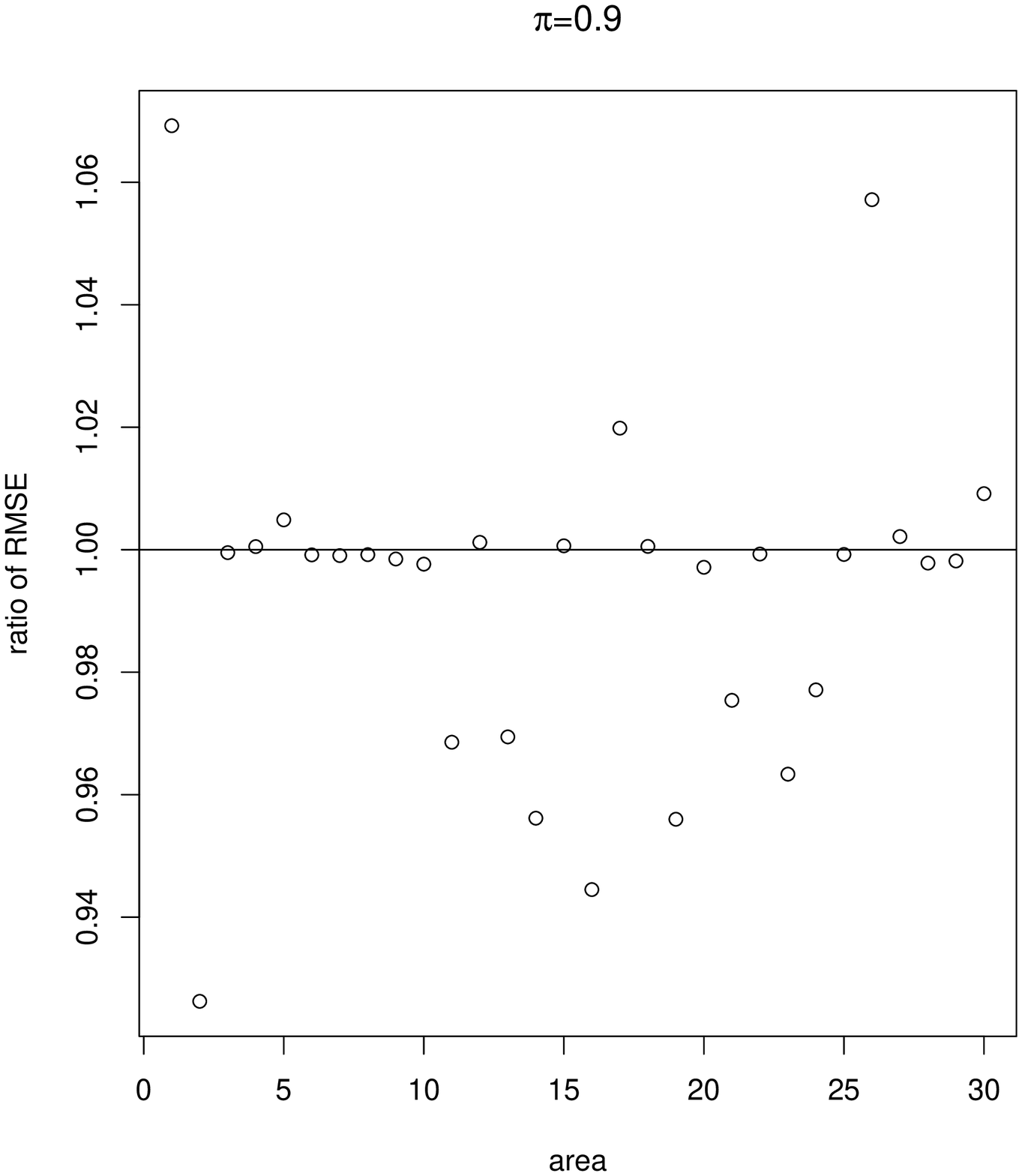}
\caption{Squared Root Mean Squared Errors of Estimation of Finite Population Mean.
\label{fig:fp}
}
\end{figure}

We next investigated the numerical performance of the small area prediction problem in the framework of a finite population.
We again used the PLP data in the Kanto region in 2001, which includes Tokyo, Kanagawa, Chiba and Saitama prefectures.
Thus the data set includes the PLP data along the Keikyu line used in the previous subsection.
The full data set we used is the land price data with covariates ($T_i$, $D_{ij}$ and $FAR_{ij}$ as used in the previous study) and each data point has its unique nearest railroad station, which we regard as a small area.
For the $i$th small area ($i=1,\dots,m$), there are $N_i$ land spots.
To consider all the observed land price data in each small area in the framework of a finite population, we analyzed only the data which belong to the small areas that have a moderately large number of data points, namely we pick up the area $i$'s with $N_i \geq 20$.
Then the resulting number of finite populations is $m=30$, and the population sizes $N_i$'s range from $20$ to $45$, but most $N_i$'s vary around $25$.
We artificially made the sampled data set and predict each finite population mean of the land price by applying UNER.
The sampling scheme is simple random sampling without replacement in each finite population and $n_i$ data are sampled in the $i$th finite population.
The sample sizes $n_i$'s are decided by some ratio $0<\pi<1$ and $100\pi$ percent of the data in each population are sampled, that is $n_i$ is the nearest integer to $N_i\times \pi$.
We considered four choices for $\pi$, namely $\pi=0.3,0.5,0.7,0.9$.
In each case, we computed the squared root mean squared errors for estimators of finite population means as
$$
{\rm SMSE}_i=\sqrt{\frac1R\sum_{r=1}^R(\muh_i^{(r)}-\mu_i)^2},
$$
where $\muh_i^{(r)}$ is the estimator of the finite population using UNER or NER, and $R=1,000$ in this study.
For both UNER and NER, we calculated $\muh_i^{(r)}$ by 5,000 posterior samples after 1,000 iterations using the method discussed in Section \ref{sec:fp}.
In the UNER estimation, the same form of the prior distribution as in the previous section was used, namely $a=5, b_1=V+2$ and $b_2=V(V+1)$ for estimated sampling error $V$.
To compare values of the SMSE for the two models, we then computed the ratio of SMSE given by ${\rm SMSE}_i^{\rm UNER}/{\rm SMSE}_i^{\rm NER}$, and provide their values in Figure \ref{fig:fp}.
It is observed from Figure \ref{fig:fp} that UNER provides better estimates than NER in some areas, but worse estimates than in several areas for the four cases of $\pi$.
Moreover, it is also revealed that an improvement of UNER over NER becomes greater as the sampling rate $\pi$ gets larger.

\section{Concluding Remarks}\label{sec:conc}
In this article, we have proposed the use of uncertain random effects in the nested error regression model called the UNER model for unit-level data.
This can be regarded as an extension of Datta and Mandal \cite{DM}.
We have used the non-informative priors for all the parameters and proposed Bayesian inferences for the linear combination of fixed effects and random effects as well as the model parameters.
We have shown that the posterior distribution is proper and the posterior variances exist under some conditions.
Through the simulation study, we have compared the UNER model with the conventional nested error regression (NER) model.
It has been revealed that UNER can provide more accurate estimates than that of NER when the underlying distribution of random effects is a mixture of a point mass on the origin and a continuous distribution.
Moreover, we have applied UNER together with NER to the PLP data and have found that the UNER model fits better than the NER model in terms of DIC.

\ \\
{\bf Acknowledgement}

\medskip
We would like to thank the associate editor and two reviewers for many valuable comments and helpful suggestions which led to an improved version of this paper.
The first author was supported in part by Grant-in-Aid for Scientific Research (15J10076) from the Japan Society for the Promotion of Science (JSPS).
Research of the second author was supported in part by Grant-in-Aid for Scientific Research  (15H01943 and 26330036) from the Japan Society for the Promotion of Science.

\appendix
\medskip

\section{Proof of Theorem \ref{thm:pos}.}
Let $\pi^{\ast}$ be the right side of (\ref{pos}).
For part (a), we shall show that 
$$
\sum_{\u\in \{0,1\}^m}\int \pi^{\ast}(\v,\u,\bbe,\si^2,\tau^2,p|D)d\v d\bbe d\si^2 d\tau^2 dp<\infty,
$$
namely the integral for each $\u$ is finite.
We first prove for the case $\u=(0,\ldots,0)^{\top}$.
In this case, the integral reduces to 
$$
\int (\si^2)^{-(N+1)/2}(1-p)^{m-1/2}
\exp\Big \{-\frac{1}{2\si^2}\sum_{i=1}^m\sum_{j=1}^{n_i}(y_{ij}-\x_{ij}^{\top}\bbe)^2\Big \}d\bbe d\si^2 d dp.
$$
It is noted that $\int p^{-1/2}(1-p)^{m-1/2}dp=B(1/2,m+1/2)$, where $B(a,b)$ is a beta function.
Then the integral is finite since the posterior distribution of the usual linear regression for the Jeffreys prior is proper if the conditions given in Theorem \ref{thm:pos} are satisfied.

For the integral in the case $z\geq 1$, using $p^{z-1/2}(1-p)^{m-z-1/2}\leq 1$, it is sufficient to show that
$$
\int \pi_{u}(\v,\si^2,\tau^2,\bbe)d\v  d\bbe d\si^2 d\tau^2<\infty,
$$
for
\begin{equation}
\pi_{u}(\v,\si^2,\tau^2,\bbe)=
\begin{cases}
\pi_{u 1}(\v,\si^2,\tau^2,\bbe) & (z>a)\\
\pi_{u 2}(\v,\si^2,\tau^2,\bbe) & (0<z\leq a)
\end{cases}
\end{equation}
where 
\begin{align*}
\pi_{u 1}(\v,\si^2,\tau^2,\bbe)
&= (\si^2)^{-(N+1)/2}(\tau^2)^{-(z+1)/2}\prod_{i=1}^m\de_0(v_i)^{1-u_i}\\
& \ \ \ \ \ \times\prod_{i=1}^m\Big [ \exp\Big (-\frac{\sum_{j=1}^{n_i}(y_{ij}-\x_{ij}^{\top}\bbe-v_i)^2}{2\si^2}-\frac{u_iv_i^2}{2\tau^2}\Big  )\Big  ],
\end{align*}
and
\begin{align*}
\pi_{u 2}(\v,\si^2,\tau^2,\bbe)
&= (\si^2)^{-(N+1)/2}(\tau^2)^{-z/2}\pi_{\ast}(\tau^2)\prod_{i=1}^m\de_0(v_i)^{1-u_i}\\
& \ \ \ \ \ \ \times\prod_{i=1}^m\Big [ \exp\Big (-\frac{\sum_{j=1}^{n_i}(y_{ij}-\x_{ij}^{\top}\bbe-v_i)^2}{2\si^2}-\frac{u_iv_i^2}{2\tau^2}\Big  )\Big  ].
\end{align*}
To show the integrability of $\pi_{u 1}$ and $\pi_{u 2}$, we consider the case of $\u$ with $\sum_{i=1}^mu_i=k$.
Without loss of generality, we assume that $u_i=1$ for $i=1,\ldots,k$ and $u_i=0$ for $i=k+1,\ldots,m$.
Then $\pi_{u 1}(\v,\si^2,\tau^2,\bbe)$ can be rewritten as
\begin{align*}
\pi_{u 1}&(\v,\si^2,\tau^2,\bbe)\\
&= (\si^2)^{-(N+1)/2}(\tau^2)^{-(k+1)/2}\prod_{i=1}^k\Big [ \exp\Big (-\frac{\sum_{j=1}^{n_i}(y_{ij}-\x_{ij}^{\top}\bbe-v_i)^2}{2\si^2}-\frac{v_i^2}{2\tau^2}\Big  )\Big  ]\\
&\ \ \ \ \times \Big [\prod_{i=k+1}^m \exp\Big (-\frac{\sum_{j=1}^{n_i}(y_{ij}-\x_{ij}^{\top}\bbe)^2}{2\si^2}\Big  )\de_0(v_i)\Big  ].
\end{align*}
We define $N$-dimensional vector $\s(\v_{\ast})=(\s_{(1)}(\v_{\ast})^{\top},\s_{(2)}^{\top})^{\top}$ as $\s_{(1)}(\v_{\ast})=((\y_1-v_i\1_{n_i})^{\top},\ldots,(\y_k-v_k\1_{n_k})^{\top})^{\top}$ and $\s_{(2)}=(\y_{k+1}^{\top},\ldots,\y_{m}^{\top})^{\top}$ for $\v_{\ast}=(v_1,\ldots,v_k)^{\top}$.
Then, if $N>q$, we have
\begin{align}
\int& \pi_{u 1}(\v,\si^2,\tau^2,\bbe)d\bbe
\label{int1}\\
&\propto (\si^2)^{-(N-q-1)/2-1}(\tau^2)^{-(k-1)/2-1}\exp\Big (-\frac{\s(\v_{\ast})'\A\s(\v_{\ast})}{2\si^2}-\frac{\sum_{i=1}^kv_i^2}{2\tau^2}\Big  ),
\nonumber
\end{align}
where $\A=\I_N-\X(\X^{\top}\X)^{-1}\X^{\top}$.
The right side is integrable with respect to $\si^2$ and $\tau^2$ since $N>q+1$ and $k\geq a>1$, whereby we obtain
$$
\int \pi_{u 1}(\v,\si^2,\tau^2,\bbe)d\bbe d\si^2d\tau^2\propto \pi_{u 1}(\v_{\ast})\prod_{i=k+1}^m\de_0(v_i),
$$
where
$$
\pi_{u 1}(\v_{\ast})=\Big \{\s(\v_{\ast})^{\top}\A\s(\v_{\ast})\Big  \}^{-(N-q-1)/2}\Big (\v_{\ast}^{\top}\v_{\ast}\Big  )^{-(k-1)/2}.
$$
In what follows, we show that $\pi_{u 1}(\v_{\ast})$ is integrable.
To this end, we note that
$$
\int_{\Re^k}\pi_{u 1}(\v_{\ast})d\v=\int_{\v_{\ast}^{\top}\v_{\ast}\leq 1}\pi_{u 1}(\v_{\ast})d\v+\int_{\v_{\ast}^{\top}\v_{\ast}\geq 1}\pi_{u 1}(\v_{\ast})d\v,
$$
and we evaluate the two integrals separately.
For the first term, since $\A$ is idempotent and ${\rm rank}(\A)=N-q\ (>0)$, there exists $c(\y)>0$ such that $\s(\v_{\ast})^{\top}\A\s(\v_{\ast})>c(\y)$ for all $\v_{\ast}$.
Then we have
\begin{align*}
\int_{\v_{\ast}^{\top}\v_{\ast}\leq 1}\pi_{u 1}(\v_{\ast})d\v&\leq c^{-(N-q-1)/2}\int_{\v_{\ast}^{\top}\v_{\ast}\leq 1}\Big (\v_{\ast}^{\top}\v_{\ast}\Big  )^{-(k-1)/2}d\v\\
&=c^{-(N-q-1)/2}V(S^k)\int_{0}^1(r^2)^{-(k-1)/2}(r^2)^{(k-1)/2}dr<\infty,
\end{align*}
where $V(S^k)$ is the volume of the unit sphere in $\Re^k$.
For the second term, it follows that 
$$
\int_{\v_{\ast}^{\top}\v_{\ast}\geq 1}\pi_{u 1}(\v_{\ast})d\v
=\int_{\v_{\ast}^{\top}\v_{\ast}\geq 1}\Big \{\s(\v_{\ast})^{\top}\A\s(\v_{\ast})\Big  \}^{-(N-q-1)/2}(\v_{\ast}^{\top}\v_{\ast})^{-(k-1)/2}d\v.
$$
Since $\s(\v_{\ast})^{\top}\A\s(\v_{\ast})$ is a quadratic function of $\v_{\ast}$, the integral is finite as far as $N>q+2$.
For the integrability of $\pi_{u _2}$, we carry out integration with respect to $\bbe, \si^2$ and $\tau^2$ to get
$$
\int \pi_{u 2}(\v,\si^2,\tau^2,\bbe)d\bbe d\si^2d\tau^2\propto \pi_{u 2}(\v_{\ast})\prod_{i=k+1}^m\de_0(v_i).
$$
Since for $N>q+1$,
\begin{align*}
\pi_{u 2}(\v_{\ast})=&
\Big \{\s(\v_{\ast})^{\top}\A\s(\v_{\ast})\Big  \}^{-(N-q-1)/2}\Big (\v_{\ast}^{\top}\v_{\ast}+2b_2\Big  )^{-k/2-b_1}\\
\leq& c^{-(N-q-1)/2}(2b_2)^{-k/2-b_1},
\end{align*}
it follows that $\pi_{u 2}(\v_{\ast})$ is integrable as far as $N>q+1$.
Thus the proof of part (a) is established.

\bigskip
For the proof of part (b), it is sufficient to show that the posterior second moments are finite.
Since the statement for $p$ is clear, we establish the result for $\bbe, \si^2$ and $\tau^2$.
As in the proof of part (a), we consider the three cases where $z>a, \ 0<z\leq a$ and $z=0$.
By replacing $N+1$ in expressions of $\pi_{u 1},\pi_{u 2}$ and $\pi_{u 3}$ with $N+5$, it follows that $\E[(\si^2)^2|D]<\infty$ when $N>q+6$.

For $\E[\bbe\bbe^{\top}|D]$, we first note that
\begin{align*}
\int_{\Re^q}&\bbe\bbe^{\top}\exp\Big (-\frac{(\s(\v_{\ast})-\X\bbe)^{\top}(\s(\v_{\ast})-\X\bbe)}{2\si^2}\Big  )d\bbe\\
=&(\si^2)^{q/2}|\X^{\top}\X|^{-1/2}\exp\Big (-\frac{\s(\v_{\ast})^{\top}\A\s(\v_{\ast})}{2\si^2}\Big  )(\X^{\top}\X)^{-1}
\\
&\times \Big \{\si^2\I_q+\X^{\top}\s(\v_{\ast})\s(\v_{\ast})^{\top}\X(\X^{\top}\X)^{-1}\Big  \}\\
=&(\si^2)^{(q+2)/2}h(\X,\s(\v_{\ast}),\si^2)
\\
&+(\si^2)^{q/2}h(\X,\s(\v_{\ast}),\si^2)\X^{\top}\s(\v_{\ast})\s(\v_{\ast})^{\top}\X(\X^{\top}\X)^{-1},
\end{align*}
for 
$$
h(\X,\s(\v_{\ast}),\si^2)=|\X^{\top}\X|^{-1/2}\exp\Big (-\frac1{2\si^2}\s(\v_{\ast})^{\top}\A\s(\v_{\ast})\Big  )(\X^{\top}\X)^{-1}.
$$
Then it follows that
\begin{align*}
\int \bbe\bbe^{\top} \pi_{u 1}&(\v,\bbe,\si^2,\tau^2)d\v d\bbe  d\si^2 d\tau^2\\
\propto& \I_q \int_{\Re^k}\Big \{\s(\v_{\ast})^{\top}\A\s(\v_{\ast})\Big  \}^{-(N-q-3)/2}\Big (\v_{\ast}^{\top}\v_{\ast}\Big  )^{-(k-1)/2}d\v
\\
&+\int_{\Re^k}\X^{\top}\s(\v_{\ast})\s(\v_{\ast})^{\top}\X\pi_{u 1}(\v_{\ast})d \v.
\end{align*}
Since $\v_{\ast}\v_{\ast}^{\top}\leq (\v_{\ast}^{\top}\v_{\ast})\I_q$, the second term is finite if $k>5$ for all $k\geq a$, namely $a>5$.
The first term is also finite as far as $N>q+4$.
For the other cases $0<z\leq a$ and $z=0$, we can similarly show that $\int \bbe\bbe^{\top} \pi_{u 2}(\v,\bbe,\si^2,\tau^2)d\v d\bbe d\si^2 d\tau^2$ and $\int \bbe\bbe^{\top} \pi_{u 3}(\v,\bbe,\si^2,\tau^2)d\v d\bbe d\si^2 d\tau^2$ are finite under the condition given in Theorem \ref{thm:pos}.

Finally, for $\E[\tau^2)^2|D]$, it follows that 
\begin{align*}
\int &(\tau^2)^2 \pi_{u 1}(\v,\bbe,\si^2,\tau^2)d\v d\bbe d\si^2 d\tau^2\\
\propto& \int(\si^2)^{-(N-p-1)/2-1}(\tau^2)^{-(k-5)/2-1}
\\
&\times \exp\Big (-\frac{\s(\v_{\ast})^{\top}\A\s(\v_{\ast})}{2\si^2}-\frac{\sum_{i=1}^kv_i^2}{2\tau^2}\Big  )d\v d\tau^2 d\si^2\\
\propto& \int \Big \{\s(\v_{\ast})^{\top}\A\s(\v_{\ast})\Big  \}^{-(N-p-1)/2}\Big (\v_{\ast}^{\top}\v_{\ast}\Big  )^{-(k-5)/2}d\v
\end{align*}
which is finite as far as $k>5$ for all $k\geq a$, namely $a>5$.
In the cases of $0<z\leq a$ and $z=0$, it is integrable if 
$$
\int_0^{\infty}(\tau^2)^2\pi_{\ast}(\tau^2)d\tau^2<\infty,
$$
which can be established since $b_1>3$.
Thus we complete the proof of part (b).

\ \\
\ \\

\end{document}